\documentclass[longauth]{aa}

\usepackage{txfonts}
\usepackage{graphicx}
\usepackage{natbib}
\bibpunct{(}{)}{;}{a}{}{,} 

\begin{document}

\newcommand{\Cerenkov}{Cherenkov\ }
\newcommand{\HMS}[3]{$#1^{\mathrm{h}}#2^{\mathrm{m}}#3^{\mathrm{s}}$}
\newcommand{\DMS}[3]{$#1^\circ #2' #3''$}
\newcommand{\TODO}[1]{\textbf{TODO: \emph{#1}}}

\newcommand{\HESSONE}{HESS~J1718$-$385}
\newcommand{\PSRONE}{PSR~J1718$-$3825}

\newcommand{\HESSTWO}{HESS~J1809$-$193}
\newcommand{\PSRTWO}{PSR~J1809$-$1917}
\newcommand{\StatSysErr}[3]{$#1 \pm #2_{\mathrm{stat}} \pm #3_{\mathrm{sys}}$}

\newcommand{\D}{\mathrm{d}}

\renewcommand{\cite}{\citep} 

\title{Discovery of two candidate pulsar wind nebulae\\
  in very-high-energy gamma rays} 
\titlerunning{Two new pulsar wind nebula candidates} 
\authorrunning{The H.E.S.S. Collaboration}

\author{F. Aharonian\inst{1,13}
 \and A.G.~Akhperjanian \inst{2}
 \and A.R.~Bazer-Bachi \inst{3}
 \and B.~Behera \inst{14}
 \and M.~Beilicke \inst{4}
 \and W.~Benbow \inst{1}
 \and D.~Berge \inst{1} \thanks{now at CERN, Geneva, Switzerland}
 \and K.~Bernl\"ohr \inst{1,5}
 \and C.~Boisson \inst{6}
 \and O.~Bolz \inst{1}
% ...
}

\institute{
  Max-Planck-Institut f\"ur Kernphysik, P.O. Box 103980, D 69029
  Heidelberg, Germany
  \and
  Yerevan Physics Institute, 2 Alikhanian Brothers St., 375036 Yerevan,
  Armenia
  \and
  Centre d'Etude Spatiale des Rayonnements, CNRS/UPS, 9 av. du Colonel Roche, BP
  4346, F-31029 Toulouse Cedex 4, France
  \and
  Universit\"at Hamburg, Institut f\"ur Experimentalphysik, Luruper Chaussee
  149, D 22761 Hamburg, Germany
  \and
  Institut f\"ur Physik, Humboldt-Universit\"at zu Berlin, Newtonstr. 15,
  D 12489 Berlin, Germany
  \and
  LUTH, UMR 8102 du CNRS, Observatoire de Paris, Section de Meudon, F-92195 Meudon Cedex,
  France
  \and
  DAPNIA/DSM/CEA, CE Saclay, F-91191
  Gif-sur-Yvette, Cedex, France
  \and
  University of Durham, Department of Physics, South Road, Durham DH1 3LE,
  U.K.
  \and
  Unit for Space Physics, North-West University, Potchefstroom 2520,
  South Africa
  \and
  Laboratoire Leprince-Ringuet, IN2P3/CNRS,
  Ecole Polytechnique, F-91128 Palaiseau, France
  \and 
  Laboratoire d'Annecy-le-Vieux de Physique des Particules, IN2P3/CNRS,
  9 Chemin de Bellevue - BP 110 F-74941 Annecy-le-Vieux Cedex, France
  \and
  APC, 11 Place Marcelin Berthelot, F-75231 Paris Cedex 05, France 
  \thanks{UMR 7164 (CNRS, Universit\'e Paris VII, CEA, Observatoire de Paris)}
  \and
  Dublin Institute for Advanced Studies, 5 Merrion Square, Dublin 2,
  Ireland
  \and
  Landessternwarte, Universit\"at Heidelberg, K\"onigstuhl, D 69117 Heidelberg, Germany
% ...
}

\date{Received / Accepted}

\offprints{svenja.carrigan@mpi-hd.mpg.de, karl.kosack@mpi-hd.mpg.de}

\abstract{We present the discovery of two very-high-energy
  $\gamma$-ray sources in an ongoing systematic search for emission
  above 100\,GeV from pulsar wind nebulae in survey data from the
  H.E.S.S. telescope array.}{Imaging Atmospheric Cherenkov Telescopes
  are ideal tools for searching for extended emission from pulsar wind
  nebulae in the very-high-energy regime. H.E.S.S., with its large
  field of view of 5$^{\circ}$ and high sensitivity, gives new
  prospects for the search for these objects.}{An ongoing
  systematic search for very-high-energy emission from energetic
  pulsars over the region of the Galactic plane between $-60^\circ < l
  < 30^\circ$, $-2^\circ < b < 2^\circ$ is performed. For the
  resulting candidates, the standard H.E.S.S. analysis was applied and
  a search for multi-wavelength counterparts was performed.}{ We
  present the discovery of two new candidate $\gamma$-ray pulsar wind
  nebulae, \HESSONE\ and \HESSTWO.}{H.E.S.S. has proven to be a
  suitable instrument for pulsar wind nebula searches.}

\maketitle
\section{Introduction}

It has long been known that pulsars can drive powerful winds of highly
relativistic particles (see e.g. \citet{Gaensler06} for a
review). Confinement of these winds leads to the formation of strong
shocks, which may accelerate particles to $\sim$PeV energies. Evidence
for high-energy electrons in pulsar wind nebulae (PWNe) is provided by
the observation of both synchrotron emission in the radio through
high-energy $\gamma$-ray regimes, and by inverse Compton radiation in
the high-energy and very-high-energy (VHE, $>100$~GeV) $\gamma$-ray
range. The synchrotron photon energy flux of the highly-relativistic
PWN electron population is proportional to both the total number of
electrons and the magnetic field energy density.  The ratio of X-ray
to $\gamma$-ray emission is then related to the magnetic field in the
nebula. Given the general difficulty of estimating the magnetic field
in such objects, it is hard to probe the spatial and energy
distributions of the accelerated particles with X-ray observations
alone. Measurements of high-energy $\gamma$-ray radiation resulting
from inverse Compton scattering have a considerable advantage in that
they provide a direct view of the parent particle population if the
target photon fields are spatially uniform and have a well-known
density, e.g where the dominant target is the cosmic microwave
background radiation (CMBR) or possibly the dense interstellar
radiation fields present in the inner Galaxy.

The best studied example of a PWN is the Crab nebula, which exhibits
strong non-thermal emission across most of the electromagnetic
spectrum from radio to $>$50~TeV
$\gamma$-rays~\cite{WHIPPLE:crab,CANGAROO:crab,HEGRA:crab}. In
addition, VHE $\gamma$-rays have been detected from other young PWNe,
one of them located within the supernova remnant
\mbox{MSH~15$-$5\emph{2}} \cite{HESS:MSH1552}. More recently, VHE
$\gamma$-ray emission has been detected from the Vela\,X PWN
\cite{HESS:velax}. In contrast to the Crab pulsar, the Vela pulsar is
an order of magnitude older ($\sim$11\,kyr) and its nebula is
significantly offset from the pulsar position, both in X-rays and VHE
$\gamma$-rays, possibly due to the expansion of the supernova blast
wave into an inhomogeneous interstellar medium
\cite{blondin01:PWN}. Offset nebulae in both X-rays and VHE
$\gamma$-rays have also been observed in the Kookaburra Complex
\cite{HESS:kookaburra} and for the PWN associated with the
$\gamma$-ray source HESS~J1825$-$137 \cite{HESS:J1825a,
HESS:J1825b}. The latter source appears much brighter and more
extended in VHE $\gamma$-rays than in keV X-rays. This suggests that
searches at TeV energies are a powerful tool for detecting PWNe.

Motivated by these detections, an ongoing systematic search for VHE
$\gamma$-ray sources associated with high spin-down energy loss rate
pulsars is being performed, using data obtained with the
H.E.S.S.~(High Energy Stereoscopic System) instrument. The
H.E.S.S.~telescope system is currently the most sensitive instrument
for $\gamma$-ray astronomy in the energy regime above 100\,GeV. The
instrument is an array of four 13\,m diameter Imaging Atmospheric
Cherenkov Telescopes (IACTs) located in the Khomas highland of
Namibia. The system has an angular resolution better than
0.1$^{\circ}$, an energy resolution of $\sim$15\,\% and a point-source
sensitivity of $<2.0\times10^{-13}$~cm$^{-2}$~s$^{-1}$ (1\,\% of the
flux from the Crab nebula, \cite{HESS:crab}) for a 5\,$\sigma$
detection in 25 hours of observation. The $5^{\circ}$ field of view of
the instrument and its Southern Hemisphere location make it ideal for
surveying the inner Galactic plane.

The VHE $\gamma$-ray data set used in the search includes all data
used in \citet{HESS:scanpaper2}, an extension of the survey to
$-60^{\circ} < l < -30^{\circ}$, dedicated observations of Galactic
targets and re-observations of H.E.S.S. survey sources. It spans
Galactic longitudes $-60^{\circ} < l < 30^{\circ}$ and Galactic
latitudes $-2^{\circ} < b < 2^{\circ}$, a region covered with high
sensitivity in the survey. These data are being searched for VHE
emission from pulsars from the Parkes Multibeam Pulsar
Survey~\cite{Parkes1}. The search for a possible $\gamma$-ray excess
is done in a circular region with radius $\theta = 0.22^{\circ}$
($\theta = 13.2'$) \citep[as in][]{HESS:scanpaper2} around each pulsar
position, sufficient to encompass a large fraction of a possible
PWN. The statistical significance of the resulting associations of the
VHE $\gamma$-ray source with the pulsar is evaluated by repeating the
procedure for randomly generated pulsar samples, modelled after the
above-mentioned parent population.

In this search, it is found that pulsars with high spin-down energy
loss rates are on a statistical basis accompanied by VHE emission. The
VHE sources described in this paper have high statistical
significances and are therefore also detected in a blind search for
VHE sources \cite[similar to that reported in][]{HESS:scanpaper2}.

The search for VHE $\gamma$-ray sources near the pulsars \PSRONE\ and
\PSRTWO\ revealed two new VHE $\gamma$-ray sources, \HESSONE\ and
\HESSTWO, respectively.  This paper deals with the results from the
HESS data analysis of these two new sources, and with their possible
associations with the pulsars and other objects seen in radio and
X-ray wavelengths. In the data set used for this search, other new VHE
sources have been detected that are not associated with pulsars, and
will be reported elsewhere.

\section{H.E.S.S. Observations and Analysis}

\begin{table*}[h]
  \begin{center}
    {\tiny
    \vspace{4mm}
    \begin{tabular}{|l|l|l|l|l|l|l|l|l|l|l|l|l|l|}
      \hline
       & \multicolumn{6}{c|}{Sky Maps} & \multicolumn{6}{c|}{Spectra}\\
      \hline
      Source & $\theta_\mathrm{M}$ & Cuts & $t_{\mathrm{live}}$ & $\langle ZA \rangle$ & $\phi$ & $E_{\mathrm{thresh}}$ &
       $\theta_\mathrm{S}$ &Cuts & $t_{\mathrm{live}}$ & $\langle ZA \rangle$ & $\phi$ & $E_{\mathrm{thresh}}$ \\

      & ($^\circ$) & & (hrs) & ($^\circ$) & ($^\circ$) & (GeV)&($^\circ$) & & (hrs) & ($^\circ$)&($^\circ$) &(GeV) \\\hline
      \HESSONE\ & 0.11 & hard & 82 & 33.9 & 1.6 & 450 & 0.2 &hard & 73 & 32.5 & 1.1 & 450\\
      \HESSTWO\ & 0.19 & hard & 25 & 20.4 & 1.7 & 350 & 0.5 &standard & 9 & 26.5&1.0&250 \\\hline 
    \end{tabular}
     }
     \caption{Data properties and analysis parameters for the two PWN
     candidates. Here, $\theta_\mathrm{M}$ and $\theta_\mathrm{S}$ are
     the on-source integration radii for the sky maps and the spectral
     analysis, respectively, $t_{\mathrm{live}}$ is the live time,
     $\langle ZA \rangle$ is the mean zenith angle, $\phi$ is the mean
     offset of the observation position from the target position, and
     $E_{\mathrm{thresh}}$ is the analysis threshold energy.}
    \label{analysistable}
  \end{center}
\end{table*}

The data on the first source, \HESSONE, are composed primarily from
dedicated observations of the supernova remnant RX~J1713.7$-$3946
(\mbox{G347.3$-$0.5}), which is located $\sim$1.6$^{\circ}$ south-west
of \HESSONE~\cite{ASCA:RXJ1713,CANGAROO:RXJ1713,HESS:RXJ1713}. The
first H.E.S.S.\ observations of the region around the second source,
\HESSTWO, were taken as part of the systematic survey of the inner
Galaxy from May to June
2004~\cite{HESS:scanpaper1,HESS:scanpaper2}. As there was a marginally
significant VHE $\gamma$-ray signal ($2\sigma$ post-trials),
re-observations of \HESSTWO\ were taken in 2004 and 2005. After
passing the H.E.S.S.\ standard data quality criteria~\cite{HESS:crab}
based on hardware and weather conditions, the data set for \HESSONE\
has a total live time of $\sim$82 hours and for \HESSTWO\ a live time
of $\sim$25 hours (see Table \ref{analysistable} for observation
properties and analysis parameters for both sources). Both data sets
come from observations whose centre positions are less than
$3^{\circ}$ offset from the respective best-fit source position. Since
most of the observations were not specifically targeted at these
sources, offsets from the pointing positions are larger than for
dedicated observations (which are typically offset by $\leq
1.0^\circ$).

The standard H.E.S.S.\ analysis scheme~\cite{HESS:crab} was applied to
the data, including optical efficiency corrections.  The off-axis
sensitivity of the system derived from Monte Carlo simulations and
observational data has been confirmed by observations of the Crab
nebula. Cuts on the scaled width and length of images (optimised on
$\gamma$-ray simulations and off-source data) were used to suppress
the hadronic background.

To produce sky maps, \emph{hard cuts} \cite{HESS:crab} were applied,
which include a rather tight cut on the shower image brightness of 200
photo-electrons and are suitable for extended, hard-spectrum sources
such as PWN. The background at each test position in the sky was
derived from a ring with a mean radius of 1~degree surrounding this
position and a width scaled to provide a background area $\sim$7 times
larger than the area of the on-source region, which is defined as a
circle of radius $\theta_\mathrm{M}$.

For spectral studies, only observations in which the camera centre is
offset by less than $2^{\circ}$ from the respective best-fit source
positions were used to reduce systematic effects due to reconstructed
$\gamma$-ray positions falling close to edge of the field of view. The
spectral significances are calculated by counting events within a
circle of radius $\theta_\mathrm{S}$ from the best fit position. The
integration radius was chosen to enclose the whole emission region for
each source while reducing systematic effects arising from morphology
assumptions. In the case of \HESSTWO, \emph{standard cuts}
\cite{HESS:crab} on the image parameters were applied, loosening the
minimum image brightness to 80 photo-electrons to extend the energy
spectrum down to $\sim$250\,GeV (at the expense of a reduced
signal-to-noise ratio), and the background was estimated from regions
with equal offset from the centre of the field of view to minimise
spectral systematic uncertainties. For \HESSONE, \emph{hard cuts} were
retained for deriving the energy spectrum as these cuts also improve
the angular resolution and therefore suppress contamination from the
nearby supernova remnant RX~J1713.7$-$3946. The proximity of this
strong source made it necessary to choose the background data from
off-source observations (matched to the zenith angle and offset
distribution of the on-source data) instead of from areas in the same
field of view. For a more detailed description of methods for
background estimation, see~\citet{HESS:bg}. The remaining live times
of the data samples were $\sim$9 hours in case of \HESSTWO\ and
$\sim$73 hours in case of \HESSONE.

\section{Results}

\subsection{\HESSONE} \label{sec:results}

The detection significance from the search for VHE $\gamma$-ray
emission within $13.2'$ of the location of \PSRONE\ is $7.9\sigma$. A
very conservative estimate of the number of trials involved
\citep[similar to][]{HESS:scanpaper2} leads to a corrected
significance of $6.2\sigma$.

\begin{figure*}[h]
  \centering
  \resizebox{0.51\hsize}{!}{\includegraphics{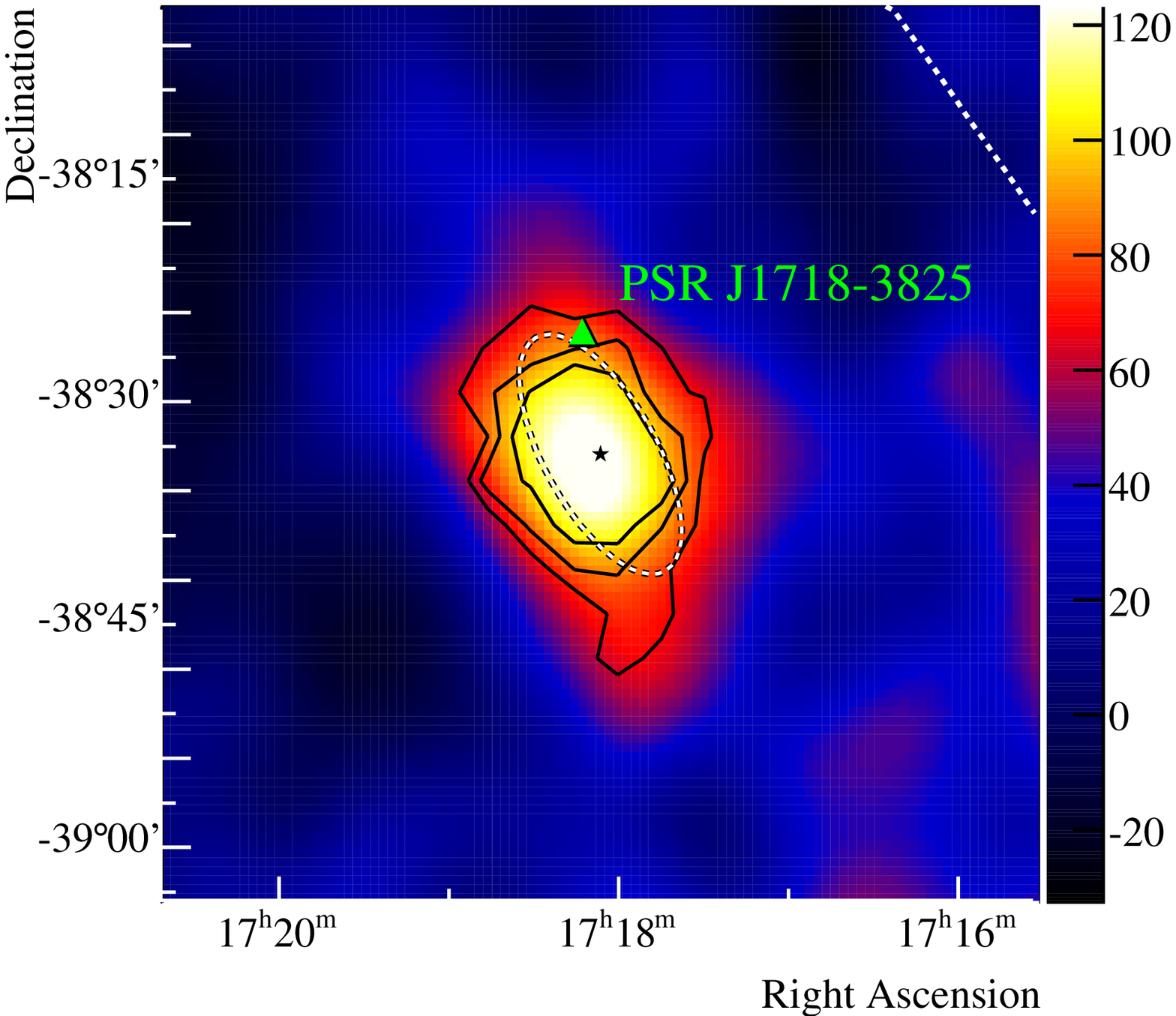}} \\
  \resizebox{0.47\hsize}{!}{\includegraphics{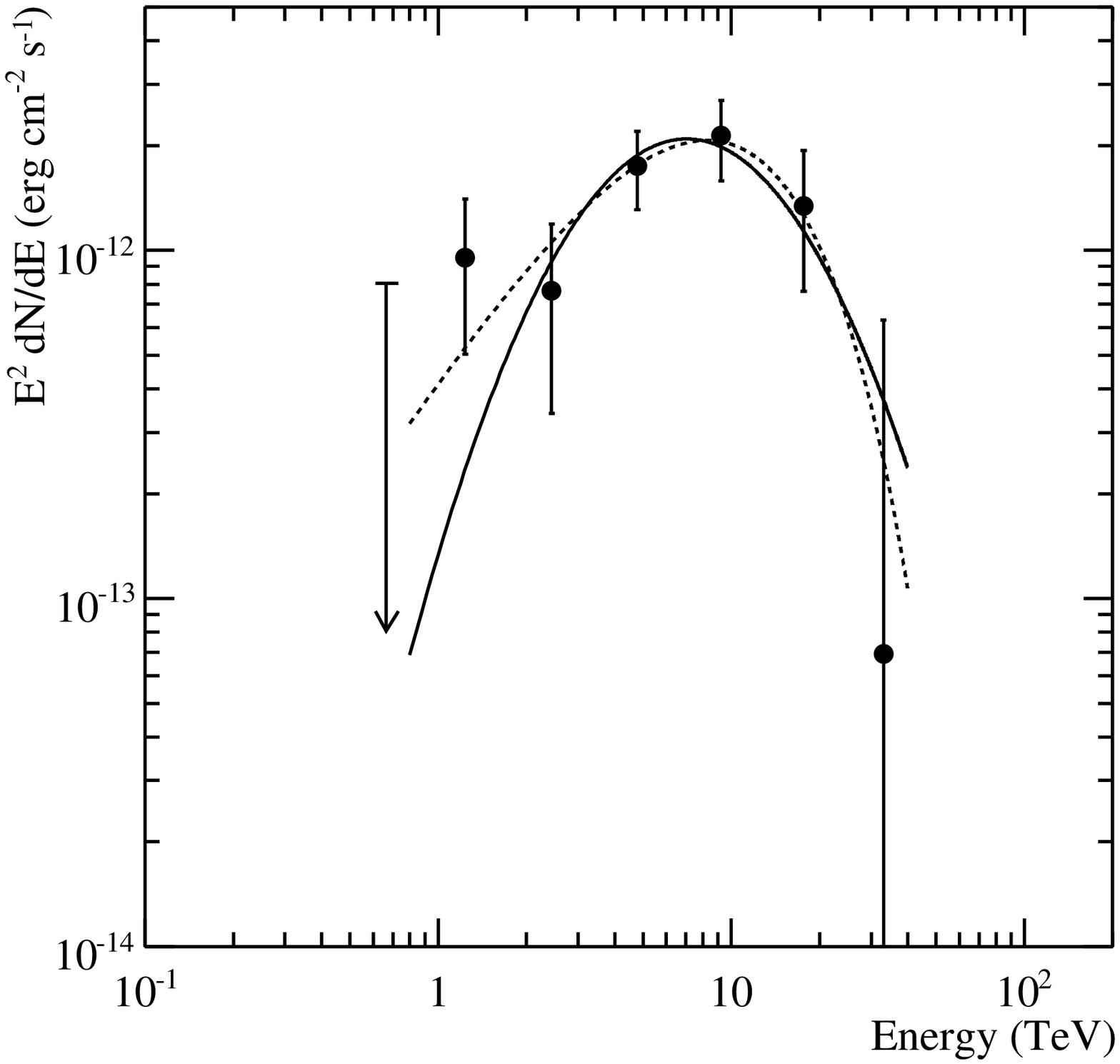}} 
  \caption{ \emph{Top:} An image of the VHE $\gamma$-ray excess counts
    of \HESSONE, smoothed with a Gaussian of width $3.8'$. The colour
    scale is set such that the blue/red transition occurs at
    approximately the 3\,$\sigma$ significance level. The black
    contours are the 4, 5 and 6\,$\sigma$ significance contours. For
    this significance calculation, events were correlated with a top
    hat function with an RMS chosen to match that of the Gaussian
    smoothing (and thus an integration radius
    $\theta_\mathrm{M}=6.6'$). The position of the pulsar \PSRONE\ is
    marked with a green triangle and the Galactic plane is shown as a
    white dotted line. The best fit position for the $\gamma$-ray
    source is marked with a black star and the fit ellipse with a
    dashed line. \emph{Bottom:} The energy spectrum of \HESSONE, which
    is fit by a curved profile (solid line). Alternatively, the fit of
    an exponentially cut-off power law is shown (dashed line, refer to
    the text for details on both fits). The first point in the
    spectrum lacks statistics due to lower exposure at small zenith
    angles and is plotted as a $2\sigma$ upper limit.}
\label{fig:J1718}
\end{figure*}

The top panel of Fig.~\ref{fig:J1718} shows the excess
count map of the $60'\times 60'$ region around \HESSONE, smoothed with
a Gaussian of width $3.8'$ chosen to reduce statistical fluctuations
while retaining source features. A two-dimensional Gaussian brightness
profile, folded with the H.E.S.S. point-spread function, is fit to the
distribution before smoothing. Its parameters are the width in two
dimensions and the position angle, defined counter-clockwise from
North. The intrinsic widths (with the effect of the point-spread
function removed) for the fit are $9' \pm 2'$ and $4' \pm 1'$
and the position angle is $\sim$33$^{\circ}$. The best fit position
for the centre of the excess is RA~=~\HMS{17}{18}{7}\,$\pm 5^{s}$,
Dec~=~$-38^{\circ}33'\pm2'$ (epoch J2000). H.E.S.S. has a systematic
pointing error of $\sim 20''$.

For the spectral analysis, a statistical significance of 6.8\,$\sigma$
(with 343 excess counts) was derived. Figure~\ref{fig:J1718} (bottom)
shows the measured spectral energy distribution for \HESSONE\ (in
$E^{2}$\,dN/dE representation). The fit of a pure power law
($dN/dE=N_0 E^{-\Gamma}$) to the spectrum gives a $\chi^2/{d.o.f.} =
11.1/4$ and a spectral index of $\Gamma=$\StatSysErr{2.1}{0.1}{0.2}.

The spectrum is fit by a curved profile (shown as the solid line):
\begin{equation}
  \frac{dN}{dE} = N_0 \left(\frac{E_{\mathrm{peak}}}{1\,\mathrm{TeV}}\right)^{-2} \left(\frac{E}{E_{\mathrm{peak}}}\right)^{(\beta \cdot \mathrm{ln}E/E_{\mathrm{peak}}) - 2.0}
\end{equation}
The peak energy $E_{\mathrm{peak}}$ is $(7 \pm 1_\mathrm{stat} \pm
1_\mathrm{sys})$\,TeV, the differential flux normalisation $N_0 = (1.3
\pm 0.3_\mathrm{stat} \pm 0.5_\mathrm{sys}) \times
10^{-12}$\,TeV$^{-1}$\,cm$^{-2}$\,s$^{-1}$ and $\beta = -0.7 \pm
0.3_\mathrm{stat} \pm 0.4_\mathrm{sys}$. This fit has a
$\chi^2/{d.o.f.} = 3.2/3$. The integral flux between $1-10$~TeV is
about 2\,\% of the flux of the Crab nebula in the same energy
range~\cite{HESS:crab}. This spectral fit was used to derive the
energy flux used later in Table~\ref{summarytable}.

Alternatively, fitting the spectrum by an exponentially cut-off power
law ($dN/dE = N_0 E^{-\Gamma} e^{-E/E_{\mathrm{cut}}}$) gives $N_0 =
(3.0 \pm 1.9_\mathrm{stat} \pm 0.9_\mathrm{sys}) \times
10^{-13}$\,TeV$^{-1}$\,cm$^{-2}$\,s$^{-1}$, photon index $\Gamma = 0.7
\pm 0.6_\mathrm{stat}\pm 0.2_\mathrm{sys}$ and a cut-off in the
spectrum at an energy of $E_{\mathrm{cut}}=(6 \pm 3_\mathrm{stat} \pm
1_\mathrm{sys})$\,TeV. This fit, which is shown as a dashed line in
the bottom panel of Fig.~\ref{fig:J1718}, has a $\chi^2/{d.o.f.} =
1.6/3$. The integral flux between $1-10$~TeV is about 2\,\% of the
flux of the Crab nebula in the same energy range~\cite{HESS:crab}.

Both the curved and exponentially cut-off power-law profiles fit the
data well; the former has the advantage of showing explicitely the
peak energy of the spectrum, which has to date only been resolved in
one other VHE source, Vela\,X \cite{HESS:velax}.

\subsection{\HESSTWO}

\begin{figure*}[h]
  \centering
  \resizebox{0.51\hsize}{!}{\includegraphics{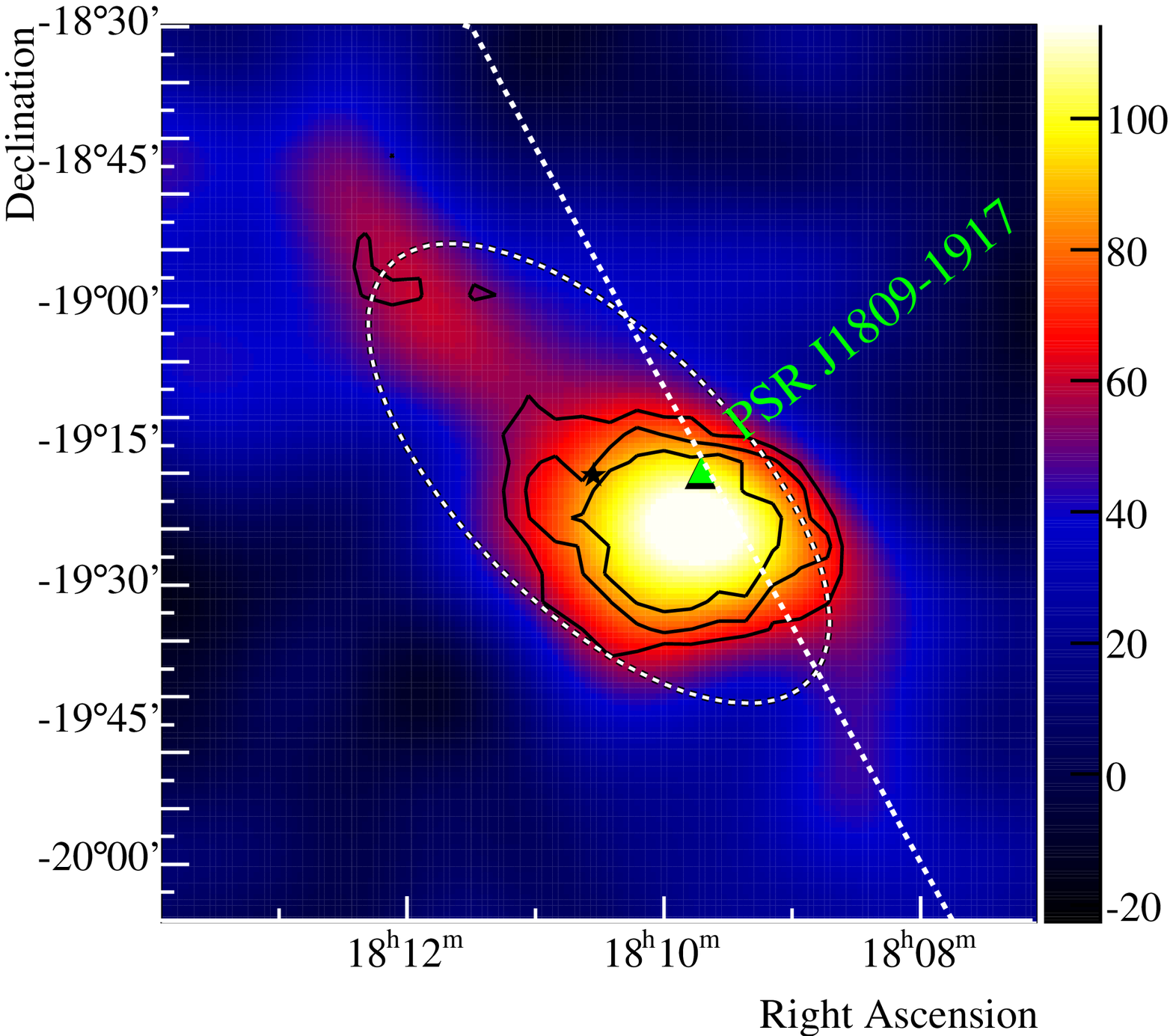}} \\
  \resizebox{0.47\hsize}{!}{\includegraphics{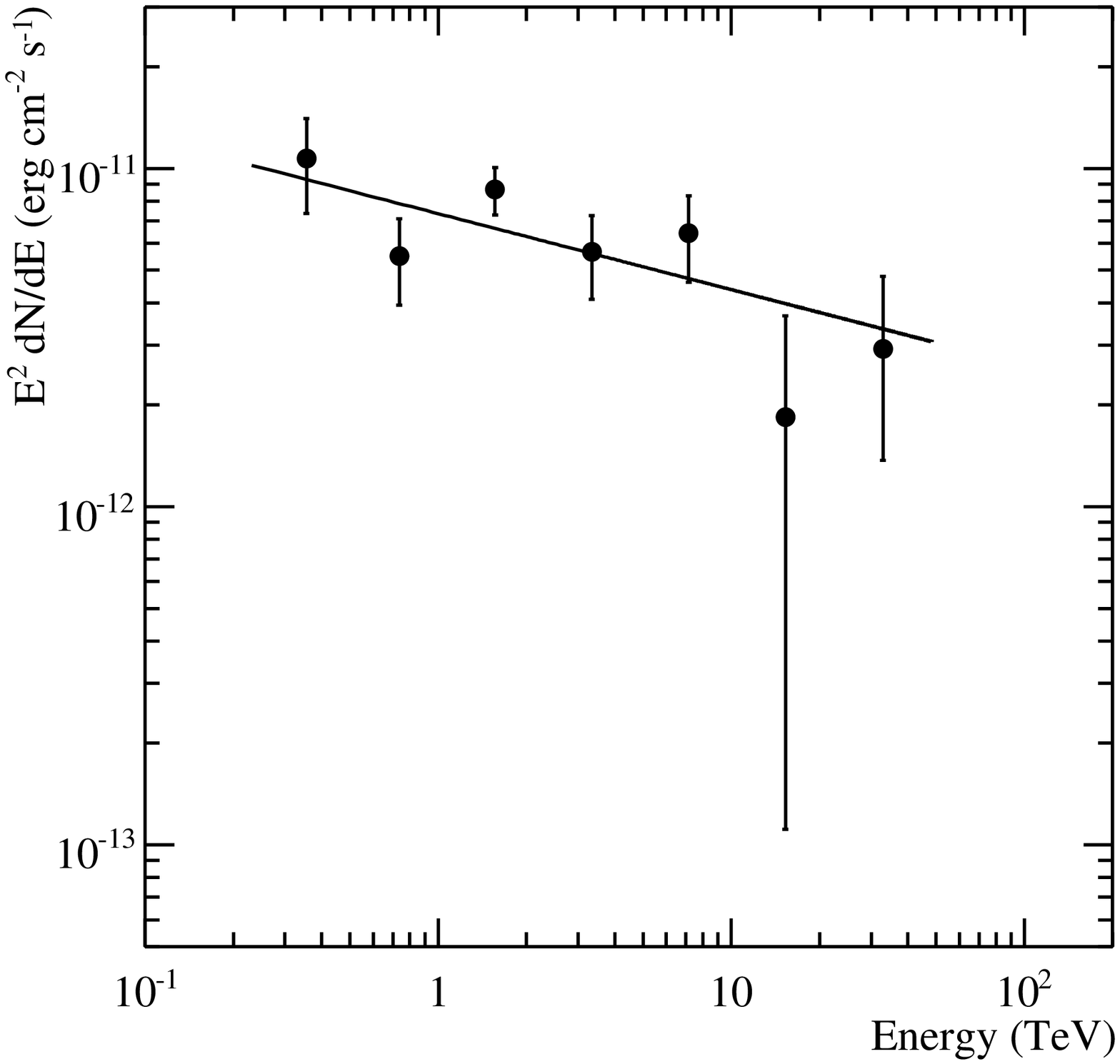}} 
  \caption{\emph{Top:} An image of the VHE $\gamma$-ray excess counts
    of \HESSTWO, smoothed with a Gaussian of width $6.6'$. The colour
    scale is set such that the blue/red transition occurs at
    approximately the 3\,$\sigma$ significance level. The black
    contours are the 4, 5 and 6\,$\sigma$ significance contours.  For
    this significance calculation, events were correlated with a top
    hat function with an RMS chosen to match that of the Gaussian
    smoothing (and thus an integration radius
    $\theta_\mathrm{M}=11.4'$). The position of the pulsar \PSRTWO\ is
    marked with a green triangle, while the Galactic plane is shown as
    a white dotted line. The best fit position for the $\gamma$-ray
    source is marked with a black star and the fit ellipse with a
    dashed line. \emph{Bottom:} The energy spectrum of \HESSTWO, which
    is fit by a power law with slope $\Gamma= 2.2\pm
    0.1_\mathrm{stat}\pm 0.2_\mathrm{sys}$.}
    \label{fig:J1809}
\end{figure*}

The detection significance from the search for VHE $\gamma$-ray
emission within $13.2'$ of the location of \PSRTWO\ is $6.8\sigma$. A
very conservative estimate of the number of trials involved \citep[as
in][]{HESS:scanpaper2} leads to a corrected significance of
$4.7\sigma$.  In Fig.~\ref{fig:J1809}, the top panel displays the
uncorrelated excess count map of the $90'\times 90'$ region around
\HESSTWO, smoothed with a Gaussian of width $6.6'$, again chosen to
reduce statistical fluctuations while retaining source features.

Again assuming a two-dimensional Gaussian brightness profile folded
with the H.E.S.S. point-spread function for this clearly extended
source, the intrinsic widths of the fit ellipse are $32' \pm 4'$ and
$15' \pm 2'$, and the position angle is $\sim$50$^{\circ}$. The best
fit position for the centre of the excess is
RA~=~\HMS{18}{10}{31}\,$\pm 12^{s}$, Dec~=~$-19^{\circ}18'\pm 2'$
(epoch J2000).

For the spectral analysis, a statistical significance of 7.6\,$\sigma$
(with 875 excess counts) was derived. The bottom panel of
Fig.~\ref{fig:J1809} shows the measured spectral energy distribution
for \HESSTWO\ (in $E^{2}$\,dN/dE representation).  The spectrum is
well fit by a power law ($dN/dE=N_0 E^{-\Gamma}$) with photon index
$\Gamma=2.2 \pm 0.1_\mathrm{stat} \pm 0.2_\mathrm{sys}$ and
differential flux normalisation at 1\,TeV $N_0=(4.6 \pm
0.6_\mathrm{stat} \pm 1.4_\mathrm{sys}) \times
10^{-12}$\,TeV$^{-1}$\,cm$^{-2}$\,s$^{-1}$ and gives a
$\chi^2/{d.o.f.}  = 6.9/5$. Its integral flux between $1-10$~TeV is
about 14\,\% of the flux of the Crab nebula in the same energy range.

\section{Possible Associations}

In an ongoing search for possible associations of energetic pulsars
with VHE emission, measured with H.E.S.S. in the Galactic plane, it is
found that both new sources presented in this paper are each plausibly
associated with such a pulsar.

The properties of both pulsars, which were discovered in the Parkes
Multibeam Pulsar Survey~\cite{Parkes1,Parkes2} are summarised in
Table~\ref{summarytable}. In this section, these probable associations
are investigated by evaluating broad-band information around the
objects.

Both \PSRONE\ and \PSRTWO\ appear to be Vela-like pulsars, as they are
of comparable age and show similar spin periods, 75\,ms and 83\,ms for
\PSRONE\ and \PSRTWO, respectively. Compared with the general sample
used in the systematic search, the associated VHE emission seems to
fit into the emerging picture of $\gamma$-ray PWNe, as implied by
sources like Vela\,X, HESS\,J1825-137 and the two PWN candidates in
the Kookaburra complex. The offset of the VHE emission from the pulsar
position is not atypical compared with other probable PWNe
associations. Such offset PWNe can be explained by either the proper
motion of the pulsar, or by interaction with non-uniform regions in
the ISM~\cite{blondin01:PWN}.

\begin{table*}[h]
  \begin{center}
    {\tiny
    \vspace{4mm}
    \begin{tabular}{|l|l|l|l|l|l|l|l|l|}
      \hline
      Pulsar & Spin-down & Spin-down & Distance & Source & 1$-$10 TeV & Size & Offset & $\epsilon_{\gamma}$  \\ 
      & luminosity & age & & & Flux & & & \\
      PSR & (erg s$^{-1}$) & (kyr) & (kpc) & HESS & (erg cm$^{-2}$s$^{-1}$) & (pc) & (pc) & (1$-$10 TeV) \\\hline 
      J1718-3825 & $1.3\times10^{36}$ & 90 & 4.2(3.6) & J1718-385 & $2.9\times10^{-12}$ & 11 & 10 & 0.5\,\%\\
      J1809-1917 & $1.8\times10^{36}$ & 51 & 3.7(3.5) & J1809-193 & $1.3\times10^{-11}$ & 35 & 13 & 1.2\,\%\\ \hline

    \end{tabular}
     }
     \caption{ \HESSONE\ and \HESSTWO\ as pulsar wind nebulae: summary
       of the properties of the possibly associated pulsars, \PSRONE\
       and \PSRTWO, respectively, taken from~\citet{ATNF}. The
       spin-down age, defined as $P/2\dot P$ (where $P$ is the period
       of the pulsar), provides an age estimate for a pulsar if the
       birth period of the pulsar was short in comparison to the
       current period and assuming a braking index $n=3$
       \cite{Kramer03:pulsar_spin}. The pulsar distances result from
       \citet{Taylor&Cordes}; a more recent estimate for the distances
       of the pulsars from~\citet{NE2001} (given in parentheses) does
       not significantly change the estimates for size, offset and
       efficiency. The size refers to the larger intrinsic width from
       the respective fits (see \S\ref{sec:results}), and the offset
       refers to the distance of the pulsar from the best fit position
       of the VHE emission region in each case. The efficiency,
       $\epsilon_\gamma$, is defined as the ratio of the $\gamma$-ray
       luminosity of the VHE source to the spin-down power of the
       pulsar.}
    \label{summarytable}
  \end{center}
\end{table*}

\begin{figure*}[h]
  \centering
  \resizebox{0.48\hsize}{!}{\includegraphics{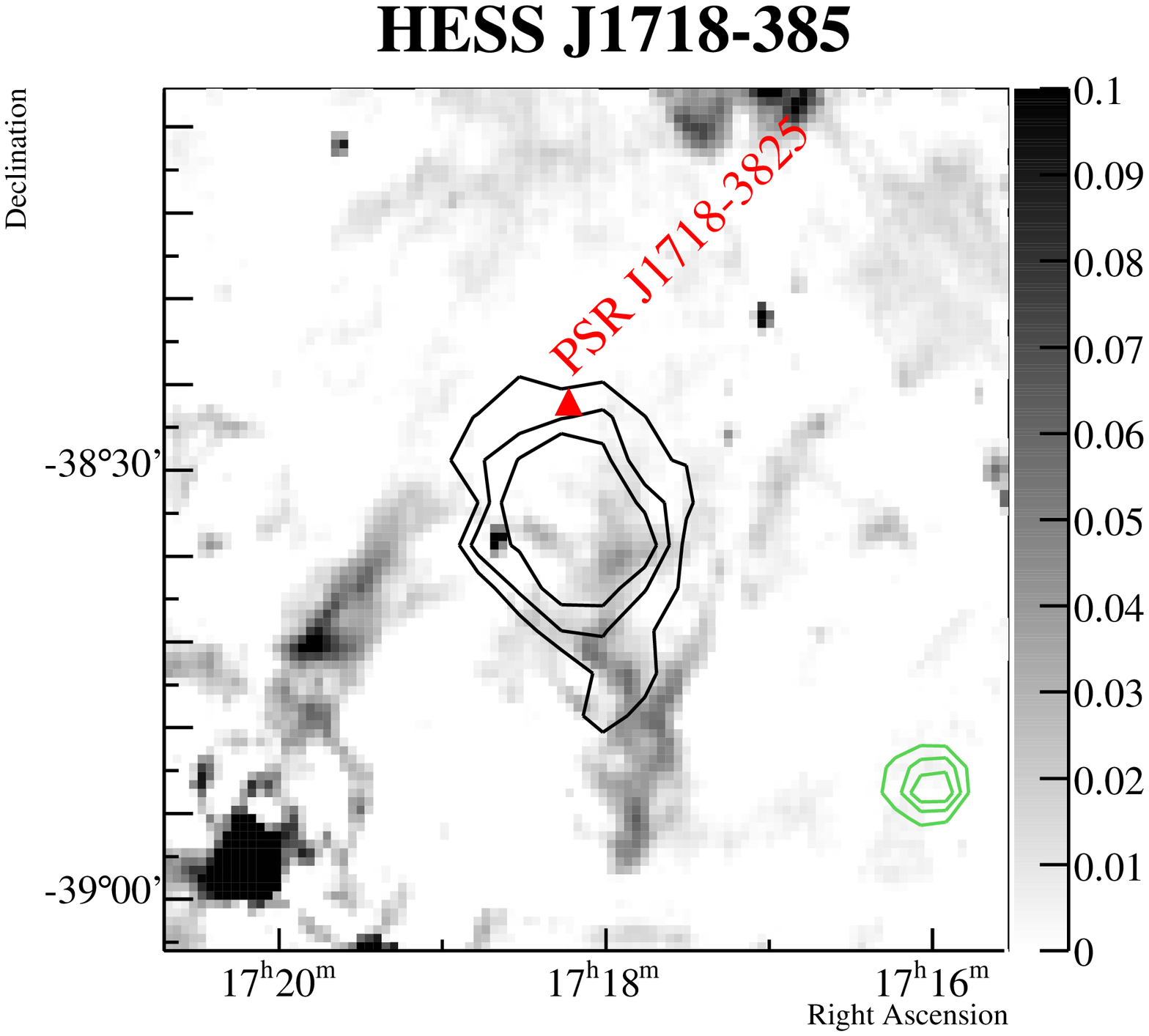}}\\ 
  \resizebox{0.48\hsize}{!}{\includegraphics{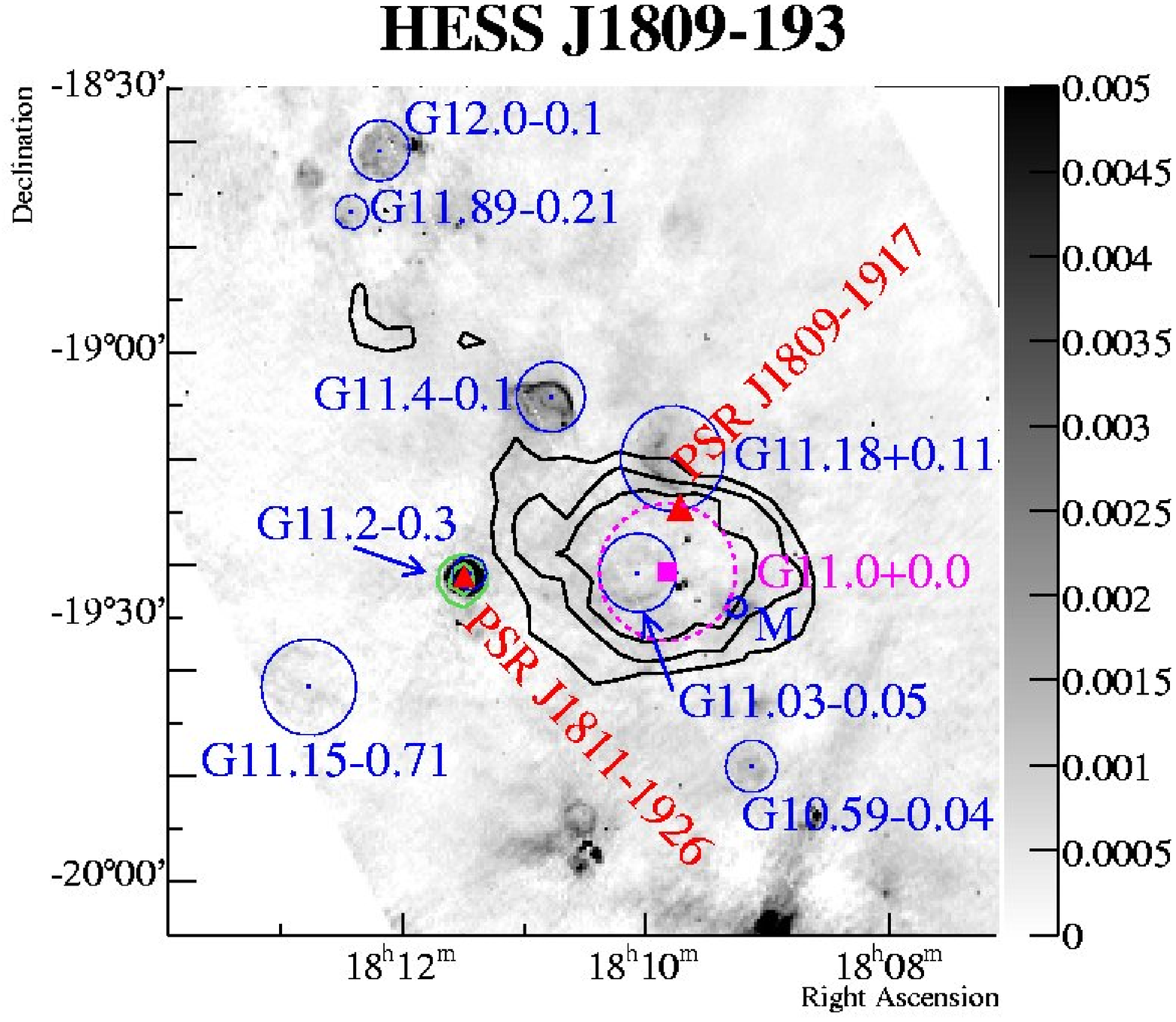}} 
  \caption{ Radio images from the Molonglo Galactic Plane Survey at
  843\,MHz~\cite{Molonglo} in the case of \HESSONE\ in the top panel,
  and from the MAGPIS survey at 1.4\,GHz~\cite{MAGPIS} in the case of
  \HESSTWO\ in the bottom panel (both in Jy/beam). The
  H.E.S.S. significance contours are overlaid in black and pulsar
  positions are marked with red triangles. Adaptively smoothed ROSAT
  hard-band X-ray contours are shown in green~\cite{ROSAT}.  Blue
  circles in the image of the \HESSTWO\ region indicate catalogued
  positions and sizes of nearby supernova remnants~\cite{green04:SNRs}
  and supernova remnant candidates~\cite{Brogan06}. The position of
  the ASCA source \mbox{G11.0$+$0.0} is marked with a magenta square
  and its size is indicated with a dashed magenta
  circle~\cite{Bamba:G11}. The most prominent source in the ROSAT data
  is the supernova remnant \mbox{G11.2$-$0.3}, which surrounds the
  X-ray emitting pulsar PSR~J1811$-$1925~\cite{Kaspi}. The MAGPIS
  supernova remnant candidate 10.8750$+$0.0875 is labeled $M$
  \cite{MAGPIS}.}
  \label{fig:X-ray}
\end{figure*}

The $\gamma$-ray source \HESSONE\ is located $\sim$0.14$^\circ$ south
of the pulsar \PSRONE\ (shown as a green triangle in
Figure\,\ref{fig:J1718} and as a red triangle in the top panel of
Figure\,\ref{fig:X-ray}). With a distance of $\sim$4\,kpc and a
spin-down flux $\dot{E}/d^2$ on the order of
$10^{35}$\,erg\,kpc$^{-2}$\,s$^{-1}$, the pulsar is energetic enough
to power \HESSONE, with an implied efficiency of conversion into 1 --
10\,TeV $\gamma$-rays of $\epsilon_\gamma = 0.5\,\%$.

As can be seen in the top panel of Fig. \ref{fig:X-ray}, no obvious
X-ray counterpart is visible for \HESSONE. There is diffuse extended
radio emission, which is partially coincident with the VHE emission,
however, this emission seems to be correlated with thermal dust
emission visible in the IRAS Sky Survey Atlas~\cite{IRAS}, suggesting
that the radio emission is thermal and is thus not likely associated
with a possible PWN. The brightest part of this diffuse feature is
catalogued as PMN~J1717$-$3846~\cite{PMN:1}. From the point of view of
positional coincidence, energetics, and lack of other counterparts,
the association of \HESSONE\ with \PSRONE\ seems plausible. To confirm
this, additional evidence from spectral and morphological studies in
VHE $\gamma$-rays and from data at other wavelengths is needed.

\PSRONE\ has so far no associated PWN detected in radio to X-ray
wavelengths. Since the VHE emission is assumed to come from inverse
Compton scattering of high-energy electrons off the CMBR, the VHE
spectrum predicts the electron spectrum and one would expect to see
X-ray synchrotron emission from this object. Under the assumption of a
magnetic field in the range of a few $\mu$G, the expected peak
synchrotron flux is $\sim 10^{-12} (B/5 \mathrm{\mu
G})^2$\,erg\,cm$^{-2}$\,s$^{-1}$. It should be possible for
instruments such as XMM-Newton to reveal an X-ray
counterpart. However, with the cut-off seen at $\sim10$\,TeV in VHE
$\gamma$-rays, the X-ray flux may peak below the XMM-Newton range.

The best fit position of the second new VHE $\gamma$-ray source,
\HESSTWO, is located at a distance of $\sim$0.2$^\circ$ east of the
pulsar \PSRTWO, while the peak of the VHE emission is at
$\sim$0.2$^\circ$ to the south of the pulsar. Its spin-down flux
$\dot{E}/d^2$, on the order of $10^{35}$~erg kpc$^{-2}$ s$^{-1}$,
implies an efficiency of $\epsilon_\gamma = 1.2\,$\%, assuming it
powers the whole emission from \HESSTWO\ (see
Table~\ref{summarytable}).

In the case of \HESSTWO, the multi-wavelength picture is much more
complicated than for \HESSONE. The bottom panel of Fig.
\ref{fig:X-ray} shows the diffuse ASCA source
\mbox{G11.0$+$0.0}~\cite{Bamba:G11,Brogan03:new_snrs,Brogan06}
coinciding with the peak of the VHE emission, which makes it a
possible X-ray counterpart to \HESSTWO. The flux of this X-ray source
is $\sim 3.8 \times 10^{-12}$~erg~cm$^{-2}$~s$^{-1}$ and its photon
index $\sim 1.6$ in the energy range 0.7 -- 10~keV. Based on its X-ray
spectrum, it was suggested that this source could be a plerionic
supernova remnant. The source distance is estimated to be 2.6~kpc,
while \PSRTWO\ is estimated to be 3.5~kpc away, however, this does not
rule out the association between the ASCA source and \PSRTWO\ as both
distance estimates suffer from large systematic uncertainties. A
further hint for the VHE emission coming from a PWN is found in public
Chandra data. There appears to be a PWN visible coincident with
\PSRTWO, that apparently shows a cometary ``tail'' structure near the
position of the pulsar, on a much smaller scale than the $\gamma$-ray
emission~\cite{CHANDRA:1809tail}. Future X-ray observations should
give more insight into its morphology.

The most prominent source in the ROSAT data is the supernova remnant
\mbox{G11.2$-$0.3}, which lies just outside the significant VHE
emission region and surrounds the X-ray emitting pulsar
PSR~J1811$-$1925 (marked with a red triangle). As the pulsar is
associated with \mbox{G11.2$-$0.3}, which is estimated to be at a
distance of $\sim 4.4$\,kpc, it is powerful enough to emit VHE
$\gamma$-rays, as its spin-down flux is then $3.3 \times 10^{35}$~erg
kpc$^{-2}$ s$^{-1}$. However, in this case, an association with the
H.E.S.S. source is highly unlikely. As Chandra has revealed, the
pulsar is very close ($\leq 8''$) to the geometric centre of the shell
and its PWN is shown to lie within the supernova
remnant~\cite{Kaspi}. Thus it seems improbable that PSR~J1811$-$1925
could have produced an extremely offset VHE PWN due to its motion, as
it seems to not yet have left its supernova remnant shell.

Finally, \PSRTWO\ might also be associated with the supernova remnant
candidates \mbox{G11.03$-$0.05},
\mbox{G11.18$+$0.11}~\cite{Brogan03:new_snrs,Brogan06}, or with the
MAGPIS supernova remnant candidate 10.8750$+$0.0875 \cite{MAGPIS},
which are all located within the VHE emission region. Given the small
angular sizes of these supernova remnants, none of them is likely to
be responsible for the bulk of the VHE emission.

\section{Summary}

\HESSONE\ might represent the first VHE $\gamma$-ray PWN found in a
systematic search for pulsar associations, despite the present lack of
a PWN detection in other wavebands. The remarkable similarity between
\HESSONE\ and other known VHE PWNe, together with the lack of other
probable counterparts, gives additional confidence. The detection of
an X-ray PWN would provide confirmation.

In the case of \HESSTWO\ the multi-wavelength picture is much more
complicated. Though the bulk of the emission can be explained by a PWN
powered by \PSRTWO, several supernova remnants as well as another PWN
may contribute to the observed VHE emission. Deeper observations in
both $\gamma$-ray and X-ray wavebands are needed to conclusively
distinguish the origin of the signal.

With the increasing number of detections of PWNe and PWN candidates
powered by high spin-down flux pulsars, VHE $\gamma$-rays are likely
to be a useful tool for discovering more of these types of objects.

\begin{acknowledgements}
  The support of the Namibian authorities and of the University of
  Namibia in facilitating the construction and operation of H.E.S.S.
  is gratefully acknowledged, as is the support by the German Ministry
  for Education and Research (BMBF), the Max Planck Society, the
  French Ministry for Research, the CNRS-IN2P3 and the Astroparticle
  Interdisciplinary Programme of the CNRS, the U.K. Particle Physics
  and Astronomy Research Council\\(PPARC), the IPNP of the Charles
  University, the South African Department of Science and Technology
  and National Research Foundation, and by the University of Namibia.
  We appreciate the excellent work of the technical support staff in
  Berlin, Durham, Hamburg, Heidelberg, Palaiseau, Paris, Saclay, and
  in Namibia in the construction and operation of the equipment.\\We
  thank M. Ostrowski for useful comments.\\We have made use of the
  ROSAT Data Archive of the Max-Planck-Institut fuer
  extraterrestrische Physik (MPE) at Garching, Germany.
\end{acknowledgements}

\bibliographystyle{aa} \bibliography{7280}

\end{document}